\documentclass[aip,jcp,amsmath,amsfonts,floatfix,reprint]{revtex4-1}
\usepackage{graphicx,epsfig}
\usepackage{color}
\usepackage[normalem]{ulem}
\usepackage{bbold}

\begin{document}
\title{Non-symmetrized quantum noise in a Kondo quantum dot}
\author{A.~Cr\'epieux}
\affiliation{Aix Marseille Univ, Universit\'e de Toulon, CNRS, CPT, 13288 Marseille, France}
\author{S.~Sahoo}
\affiliation{Univ. Grenoble Alpes, CEA, INAC, PHELIQS, F-38000 Grenoble, France}
\affiliation{Physics Department and Research Center OPTIMAS, University of Kaiserslautern, 67663 Kaiserslautern, Germany}
\author{T.Q.~Duong}
\affiliation{Aix Marseille Univ, Universit\'e de Toulon, CNRS, CPT, 13288 Marseille, France}
\author{R.~Zamoum}
\affiliation{Facult\'e des sciences et des sciences appliqu\'ees, Universit\'e de Bouira, 10000 Bouira, Algeria}
\author{M.~Lavagna}
\email{mireille.lavagna@cea.fr}
\affiliation{Univ. Grenoble Alpes, CEA, INAC, PHELIQS, F-38000 Grenoble, France}
\affiliation{Centre National de la Recherche Scientifique CNRS, France} 

\parindent = 0pt

\begin{abstract}
The fluctuations of electrical current provide information on the dynamics of electrons in quantum
devices. Understanding the nature of these fluctuations in a quantum dot is thus a crucial step insofar as this system
is the elementary brick of quantum circuits. In this context, we develop a theory for calculating the quantum noise at finite frequency in a quantum dot connected to two reservoirs in the presence of interactions and for any symmetry of the
couplings to the reservoirs. This theory is developed in the framework of the Keldysh non-equilibrium Green function technique. We establish an analytical expression for the quantum noise in terms of the various transmission amplitudes between the reservoirs 
and of some effective transmission coefficient which we define. We then study the noise as a function of the dot energy level and the bias voltage. The effects of both Coulomb interactions in the dot and asymmetric couplings with the reservoirs are 
characterized.
\end{abstract}

\maketitle

%%%%%%%%%%%%%%%%%%%%%%%%%%%%%%%%%%%%%%%%%%%%%%%%%%%%%%%%%%%%%%%%%%
%																 %
%																 %
%		INTRODUCTION											 %
%																 %
%																 %
%%%%%%%%%%%%%%%%%%%%%%%%%%%%%%%%%%%%%%%%%%%%%%%%%%%%%%%%%%%%%%%%%%

\section{Introduction}
The understanding of noise in quantum systems is a fundamental issue when one wants to control the transfer of charges in an accurate way. The efforts in that direction in the last ten years are numerous both from the experimental side
\cite{Zarchin2008,Kung2009,Basset2010,Basset2012,Ferrier2017,Fevrier2017,Delagrange2018} and from the theoretical side
\cite{Dong2008,Vitushinsky2008,Mora2008,Souza2008,Mora2009,Moca2011,Hammer2011,Muller2013,Moca2014,Crepieux2017,Stadler2018}. Some of the main issues raised by the works on noise in quantum systems are the following: (i)~Is the measured 
noise the symmetrized one or the non-symmetrized one? (ii)~Can we have over bias noise at low temperature? (iii)~How is the noise affected by the presence of Coulomb interactions? (iv)~Does the asymmetry in the couplings between the system and the 
reservoirs change the noise? The answer to the first point is known: the measured noise will be the symmetrized noise for active (classical) detector whereas it will be the non-symmetrized one for passive detector
\cite{Lesovik1997,Aguado2000,Gavish2000,Gabelli2009}. The second point is the subject of several studies\cite{Schull2009,Kaasbjerg2015,Xu2016}. To answer to the third and fourth points, we develop a theory for calculating the noise at finite frequency in a 
quantum dot (QD) coupled to two reservoirs, in the presence of Coulomb interactions in the dot and asymmetry in the couplings to the reservoirs. By using the Keldysh non-equilibrium Green function technique, we establish an analytical expression for the noise 
in terms of the transmission amplitudes between the reservoirs and of some effective transmission coefficients which will be defined. The result that we obtain for the noise can be considered as the analog of the Meir-Wingreen formula\cite{Meir1992} for the 
current. Moreover, a physical interpretation is given on the basis of the transmission of one electron-hole pair to one of the reservoirs, where it emits an energy corresponding to the measurement frequency after recombination. The results for the noise as a 
function of the dot energy level and voltage show a zero value until $|eV|=h\nu$, where $\nu$ is the frequency, followed by a signal which strongly depends on the presence of Coulomb interactions in the dot and on the asymmetry of the couplings to the 
reservoirs. These findings are compared to measurements recently performed in a Kondo carbon nanotube QD\cite{Basset2012,Delagrange2018}.

%%%%%%%%%%%%%%%%%%%%%%%%%%%%%%%%%%%%%%%%%%%%%%%%%%%%%%%%%%%%%%%%%%
%																 %
%																 %
%		NOISE										 %
%																 %
%																 %
%%%%%%%%%%%%%%%%%%%%%%%%%%%%%%%%%%%%%%%%%%%%%%%%%%%%%%%%%%%%%%%%%%

\begin{figure}[!h]
\begin{center}
\includegraphics[width=7cm]{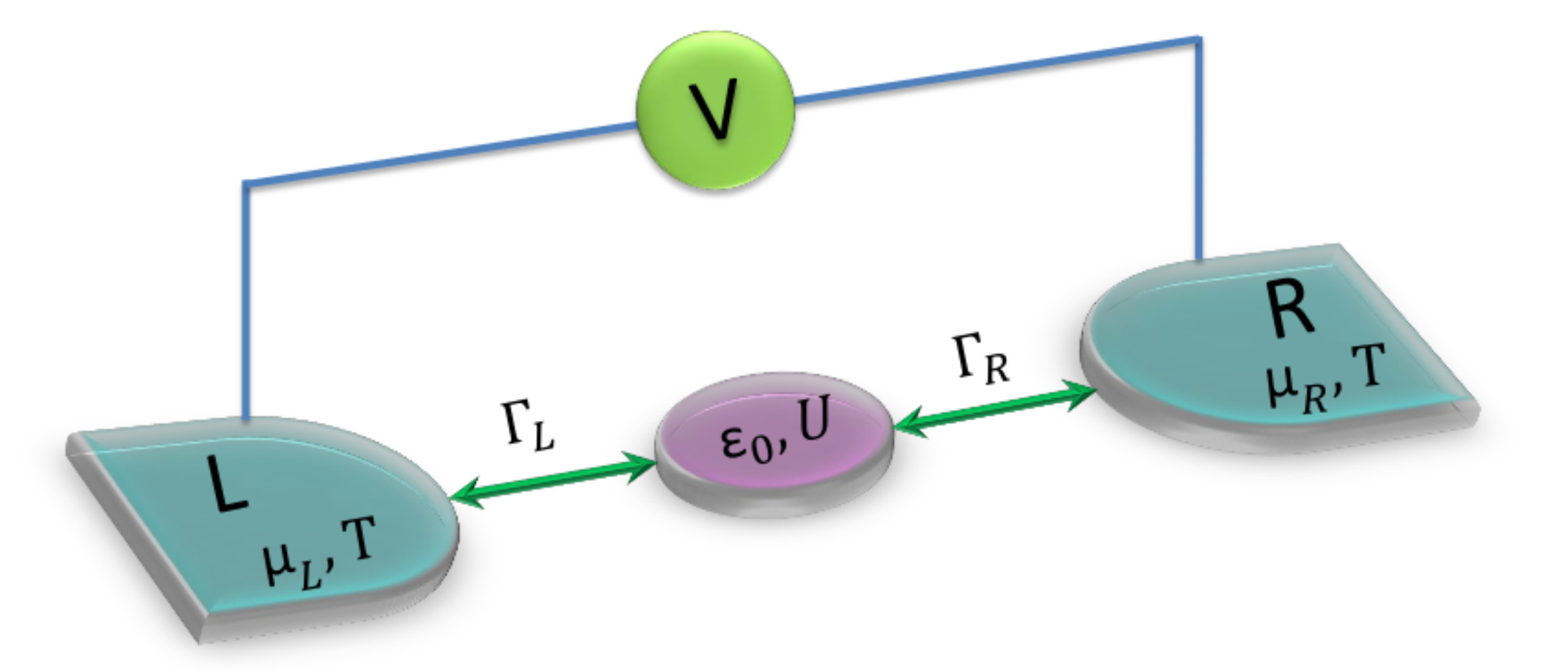}
\caption{Schematic view of the single level QD (in purple) coupled to biased reservoirs (in blue). $\mu_{L,R}$ are the chemical potentials of the reservoirs with $eV=\mu_L-\mu_R$, and $T$ their temperature. The dot is characterized by its level energy $
\varepsilon_0$ and Coulomb energy $U$. The coupling energies to the reservoirs, $\Gamma_{L,R}$, can be distinct as observed in many experiments.}
\label{schema}
\end{center} 
\end{figure}

\section{Finite-frequency noise}
We consider a single level interacting QD coupled to a left ($L$) and a right ($R$) reservoirs as depicted on Fig.~\ref{schema}. The couplings between the QD and the reservoirs are denoted $\Gamma_{L,R}$ and can be arbitrary. The asymmetry factor is 
defined as $a=\Gamma_L/\Gamma_R$. When the QD is in a steady state and the flat wideband limit is considered, we show that the finite-frequency non-symmetrized noise is given by the expression\cite{Crepieux2018}:
\begin{eqnarray}\label{NS_noise}
\mathcal{S}_{\alpha\beta}(\nu)=\frac{e^2}{h}\sum_{\gamma\delta}\int_{-\infty}^{\infty}d\varepsilon M_{\alpha\beta}^{\gamma\delta}(\varepsilon, \nu)f^e_\gamma(\varepsilon)f^h_\delta(\varepsilon-h\nu)~,
\end{eqnarray}
where $f^e_\gamma(\varepsilon)=1/(1+\exp(\varepsilon-\mu_\gamma)/k_BT)$ is the distribution function for electrons with energy $\varepsilon$ in the $\gamma$ reservoir, and $f^h_\delta(\varepsilon-h\nu)=1-f^e_\delta(\varepsilon-h\nu)$, the distribution 
function for holes with energy $\varepsilon-h\nu$ in the $\delta$ reservoir. The indices  $\alpha$, $\beta$, $\gamma$ and $\delta$ can take either the $L$ value when it relates to the left reservoir, or the $R$ value when it relates to the right reservoir. The 
expressions for the matrix elements entering in Eq.~(\ref{NS_noise}) and denoted as $M_{\alpha\beta}^{\gamma\delta}(\varepsilon, \nu)$ are given in Tab.~\ref{Mmatrix}. They depend on the transmission amplitudes $t_{\alpha\beta}(\varepsilon)$, the reflexion 
amplitudes $r_{\alpha\alpha}(\varepsilon)$, the transmission coefficients $\mathcal{T}_{\alpha\beta}(\varepsilon)$, and some effective transmission coefficients $\mathcal{T}_{LR}^{\text{eff},\alpha}
(\varepsilon)$, which are defined as:
\begin{eqnarray}\label{tdef}
t_{\alpha\beta}(\varepsilon)&=&i\sqrt{\Gamma_\alpha\Gamma_\beta}G^r(\varepsilon)~,\\
\label{rdef}
r_{\alpha\alpha}(\varepsilon)&=&1-t_{\alpha\alpha}(\varepsilon)~,\\
\label{Tdef}
\mathcal{T}
_{\alpha\beta}(\varepsilon)&=&|t_{\alpha\beta}(\varepsilon)|^2~,\\
\label{Teffdef}
 \mathcal{T}_{LR}^{\text{eff},\alpha}
(\varepsilon)&=&2\text{Re}\{t_{\alpha\alpha}(\varepsilon)\}-\mathcal{T}_{\alpha\alpha}(\varepsilon)~,
\end{eqnarray}
where $G^r(\varepsilon)$ is the retarded Green function in the QD,  and $\Gamma_\alpha$ is the coupling strength between the QD and the $\alpha$ reservoir. Eq.~(\ref{NS_noise}) is obtained considering the approximation in which the two-particle Green 
function in the dot is factorized into a product of two single-particle Green functions in the dot. From Eqs.~(\ref{tdef}-\ref{Teffdef}), we see that once $G^r(\varepsilon)$ is known, the transmission amplitudes and coefficients are entirely determined, and 
consequently, the noise given by Eq.~(\ref{NS_noise}) can be calculated explicitly. We want to underline that the effective transmission coefficient defined in Eq.~(\ref{Teffdef}) takes into account the inelastic scattering contributions\cite{Zarand2004,Borda2007}. 
When only elastic scattering is present or/and for a non-interacting system, $ \mathcal{T}_{LR}^{\text{eff},\alpha}(\varepsilon)$ coincides with $\mathcal{T}_{LR}(\varepsilon)$ since in that case, we have: $2\text{Re}\{t_{\alpha\alpha}(\varepsilon)\}=\mathcal{T}
_{\alpha\alpha}(\varepsilon)+\mathcal{T}_{LR}(\varepsilon)$, thanks to the optical theorem.

According to Eq.~(\ref{NS_noise}), $\mathcal{S}_{\alpha\beta}(\nu)$ is given by the summation over $\varepsilon$ and all possible configurations $\{\gamma,\delta\}$, of the transmission element $M_{\alpha\beta}^{\gamma\delta}(\varepsilon, \nu)$ weighted by 
the factor $f^e_\gamma(\varepsilon)f^h_\delta(\varepsilon-h\nu)$ corresponding to the probability of having a pair formed by an electron of energy $\varepsilon$  in the $\gamma$ reservoir and a hole of energy $\varepsilon-h\nu$ in the $\delta$  reservoir. Hence 
we interpret the auto-correlator $\mathcal{S}_{\alpha\alpha}(\nu)$ as the probability of transmission of an electron-hole pair from all possible configurations, to the final state for which both electron and hole are in the $\alpha$ reservoir, where by recombining it 
emits an energy $h\nu$. The additional presence of inelastic scattering does not affect this interpretation\cite{Crepieux2018}. In the case when there are several possible transmission paths, as happens for $M_{\alpha\alpha}^{\alpha\alpha}(\varepsilon, \nu)$, we 
point out the importance of considering the quantum superposition of the transmission amplitudes for all possible transmission paths\cite{Zamoum2016}.

\begin{widetext}

\begin{table}
\begin{center}
\begin{tabular}{|c||c|c|c|c|}
\hline
$M_{\alpha\beta}^{\gamma\delta}(\varepsilon,\nu)$& $\gamma=\delta=L$& $\gamma=\delta=R$&$\gamma=L$, $\delta=R$&$\gamma=R$, $\delta=L$\\ \hline\hline
$\alpha=L$&$\mathcal{T}_{LR}^{\text{eff},L}(\varepsilon)\mathcal{T}_{LR}^{\text{eff},L}(\varepsilon-h\nu)$& $\mathcal{T}_{LR}(\varepsilon)\mathcal{T}_{LR}(\varepsilon-h\nu)$ & $[1-\mathcal{T}_{LR}^{\text{eff},L}(\varepsilon)]\mathcal{T}_{LR}(\varepsilon-h\nu)$ 
& $\mathcal{T}_{LR}(\varepsilon)[1-\mathcal{T}_{LR}^{\text{eff},L}(\varepsilon-h\nu)]$\\
$\beta=L$&$+|t_{LL}(\varepsilon)-t_{LL}(\varepsilon-h\nu)|^2$&&&\\
 \hline
$\alpha=R$& $\mathcal{T}_{LR}(\varepsilon)\mathcal{T}_{LR}(\varepsilon-h\nu)$ &$\mathcal{T}_{LR}^{\text{eff},R}(\varepsilon)\mathcal{T}_{LR}^{\text{eff},R}(\varepsilon-h\nu)$ & $\mathcal{T}_{LR}(\varepsilon)[1-\mathcal{T}_{LR}^{\text{eff},R}(\varepsilon-h\nu)]
$ & $[1-\mathcal{T}_{LR}^{\text{eff},R}(\varepsilon)]\mathcal{T}_{LR}(\varepsilon-h\nu)$ \\
$\beta=R$&&$+|t_{RR}(\varepsilon)-t_{RR}(\varepsilon-h\nu)|^2$&&\\
 \hline
$\alpha=L$& $t_{LR}(\varepsilon)t^*_{LR}(\varepsilon - h\nu)$ & $t^*_{LR}(\varepsilon)t_{LR}(\varepsilon - h\nu) $  & $t_{LR}(\varepsilon)t_{LR}(\varepsilon-h\nu)$ &$t_{LR}^*(\varepsilon)t_{LR}^*(\varepsilon-h\nu)$\\
$\beta=R$&$\times[r^*_{LL}(\varepsilon)r_{LL}(\varepsilon-h\nu)-1]$&$\times[r_{RR}(\varepsilon)r^*_{RR}(\varepsilon-h\nu)-1]$&$\times r_{LL}^*(\varepsilon)r_{RR}^*(\varepsilon-h\nu)$& $\times r_{RR}(\varepsilon)r_{LL}(\varepsilon-h\nu)$\\
 \hline
$\alpha=R$& $t^*_{LR}(\varepsilon)t_{LR}(\varepsilon - h\nu)$&$t_{LR}(\varepsilon)t^*_{LR}(\varepsilon - h\nu)$  &$t_{LR}^*(\varepsilon)t_{LR}^*(\varepsilon-h\nu)$&
$t_{LR}(\varepsilon)t_{LR}(\varepsilon-h\nu)$ \\
$\beta=L$&$\times[r_{LL}(\varepsilon)r^*_{LL}(\varepsilon-h\nu)-1]$&$\times[r^*_{RR}(\varepsilon)r_{RR}(\varepsilon-h\nu)-1]$&$\times r_{LL}(\varepsilon)r_{RR}(\varepsilon-h\nu)$& $\times r^*_{RR}(\varepsilon)r^*_{LL}(\varepsilon-h\nu)$\\
 \hline
\end{tabular}
\caption{Expressions of the matrix elements $M_{\alpha\beta}^{\gamma\delta}(\varepsilon, \nu)$ involved in the Eq.~(\ref{NS_noise}) for the noise $S_{\alpha\beta}(\nu)$ of an interacting QD with arbitrary coupling symmetry to the reservoirs.}
\label{Mmatrix}
\end{center}
\end{table}

 \begin{figure}
\begin{center}
\includegraphics[width=15cm]{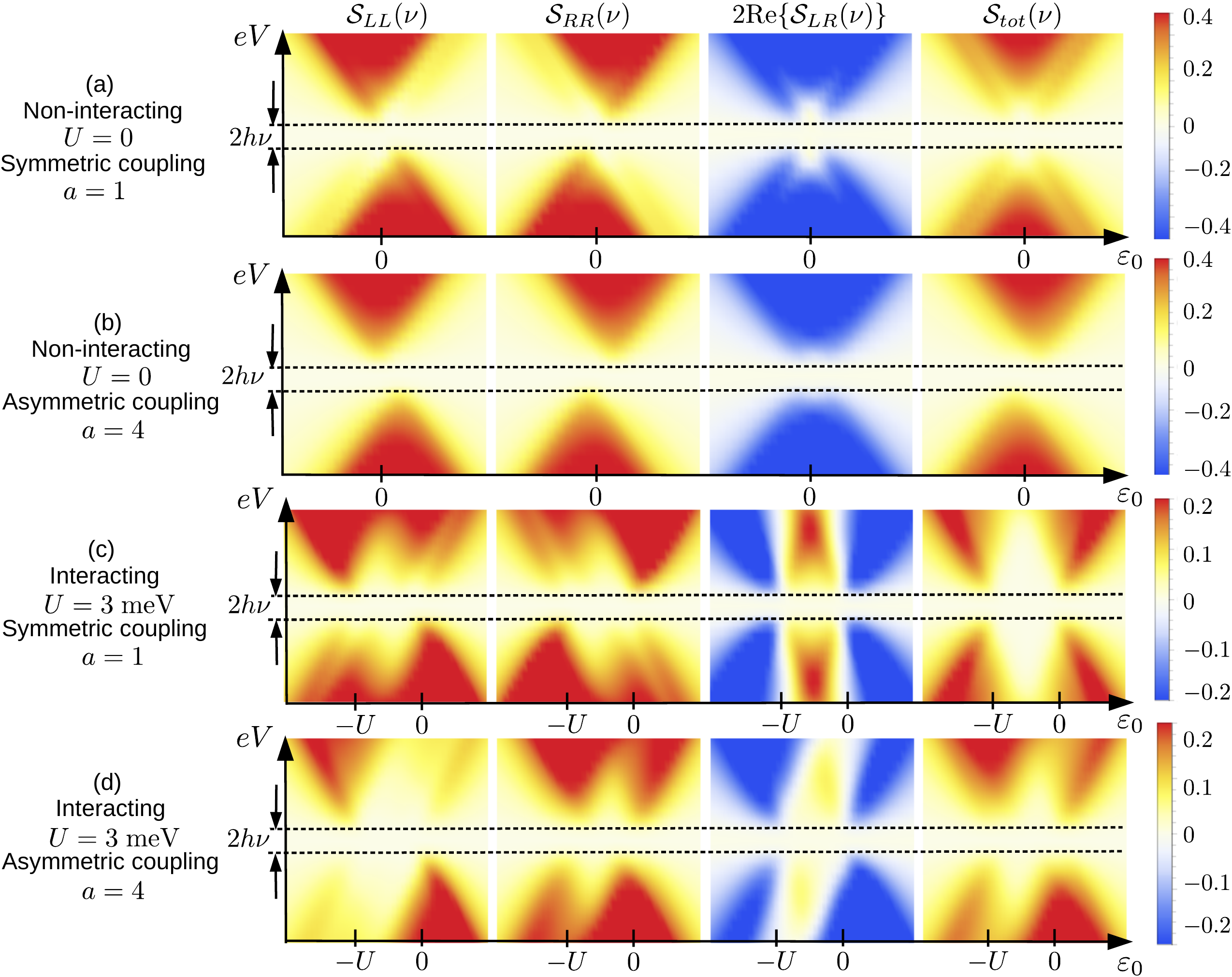}
\caption{Color-scale plots of $\mathcal{S}_{LL}(\nu)$,  $\mathcal{S}_{RR}(\nu)$, $2\text{Re}\{\mathcal{S}_{LR}(\nu)\}$ and $\mathcal{S}_{tot}(\nu)$ as a function of the dot energy level $\varepsilon_0$ (horizontal axis) and the bias voltage $eV$ (vertical axis) at 
frequency $\nu=78$ GHz and temperature $T=80$ mK. The chemical potentials are taken symmetrical: $\mu_L=eV/2$, $\mu_R=-eV/2$ and the interval of variation for the bias voltage $eV$ is [$-4$ meV$,4$ meV]. The two top rows are obtained for a non-
interacting QD ($U=0$) within the interval $\varepsilon_0\in[-3$ meV$,3$ meV$]$ whereas the two bottom rows corresponds to an interacting QD ($U=3$ meV) within the interval $\varepsilon_0\in[-6$ meV$,3$ meV$]$.}
\label{figure}
\end{center} 
\end{figure}

\end{widetext}

%%%%%%%%%%%%%%%%%%%%%%%%%%%%%%%%%%%%%%%%%%%%%%%%%%%%%%%%%%%%%%%%%%
%																 %
%																 %
%		RESULTS										 %
%																 %
%																 %
%%%%%%%%%%%%%%%%%%%%%%%%%%%%%%%%%%%%%%%%%%%%%%%%%%%%%%%%%%%%%%%%%%

\section{Kondo quantum dot}

The retarded Green function $G^r(\varepsilon)$ for the interacting single level QD is determined numerically by using a self-consistent renormalized equation-of-motion approach\cite{Roermund2010,Lavagna2015,Lavagna2018}, which applies to both 
equilibrium and non-equilibrium situations. Note that in the presence of interactions, i.e. when $U\ne 0$,  $G^r(\varepsilon)$ depends on the chemical potential $\mu_L$ and $\mu_R$. When one incorporates the expression of the Green function into 
Eqs.~(\ref{NS_noise}-\ref{Teffdef}), we are able to calculate both the auto-correlators  $\mathcal{S}_{LL}(\nu)$ and  $\mathcal{S}_{RR}(\nu)$, the cross-correlators  $\mathcal{S}_{LR}(\nu)$ and  $\mathcal{S}_{RL}(\nu)$, and the ``total'' noise defined as
\begin{eqnarray}
\mathcal{S}_\mathrm{tot}(\nu)=\frac{\mathcal{S}_{LL}(\nu)+a^2\mathcal{S}_{RR}(\nu)-a[\mathcal{S}_{LR}(\nu)+\mathcal{S}_{RL}(\nu)]}{(1+a)^2}~.\nonumber\\
\end{eqnarray}
This total noise corresponds to the noise which is measured in experiments\cite{Aguado2004,Marcos2010,Droste2015}. In Fig.~\ref{figure}, we report the color-scale plots of $\mathcal{S}_{LL}(\nu)$, $\mathcal{S}_{RR}(\nu)$, $2\text{Re}\{\mathcal{S}_{LR}(\nu)\}
=\mathcal{S}_{LR}(\nu)+\mathcal{S}_{RL}(\nu)$ and $\mathcal{S}_{tot}(\nu)$ as a function of both dot energy $\varepsilon_0$ and voltage $V$ for four sets of parameters: (a)~$U=0$ and $a=1$, (b)~$U=0$ and $a=4$, (c)~$U=3$ meV and $a=1$, and (d)~
$U=3$ meV and $a=4$.  We underline that with our choice of parameters, the estimation of the Kondo temperature with the help of the Haldane formula $k_BT_K\approx \sqrt{U\Gamma/2}\exp(\pi\varepsilon_0(\varepsilon_0+U)/2U\Gamma)$ gives $T_K\approx 
4.38$ K, which is much larger than the temperature in the reservoirs ($T=80$ mK) and larger than the frequency ($\nu=78$ GHz $\approx 3.74$  K), which ensures the QD to be in the Kondo regime when $U=3$ meV. 

We remark first that at voltage smaller in absolute value than the frequency, here $\nu=78$ GHz ($\approx 0.32$ meV), the noise is equal to zero in all graphs, as expected at low temperature (here $T=80$ mK) for the reason that the system cannot emit energy 
at a frequency larger than the energy $|eV|$ provided to it. Thus, there is a central region of width equal to $2h\nu$ (delimited by two parallel horizontal dashed lines) in the $\{\varepsilon_0,eV\}$ plane inside which the noise is strongly suppressed, in full 
agreement with experiments performed in a carbon nanotube Kondo QD\cite{Basset2012,Delagrange2018}. 

Next, we turn our interest to the effect of interactions on the dependence of the cross-correlator $\mathcal{S}_{LR}(\nu)$ with $\varepsilon_0$ and $V$. We note that when interactions are present ($U\ne 0$), the real part of the cross-correlator changes its sign 
from negative sign (blue regions) to positive sign (yellow-red regions) when $\varepsilon_0$ varies (see the third column in Figs.~\ref{figure}(c) and \ref{figure}(d)). This is not the case in the absence of interactions (see the third column in Figs.~\ref{figure}(a) and 
\ref{figure}(b)). Indeed, in that case the real part of the cross-correlator stays negative (blue) as expected for carriers (here electrons) obeying a fermionic statistic. It means that when interactions are absent, the statistic of the carriers is fermionic whereas in the 
presence of interactions, the statistic of the carriers looks bosonic-like in some regions and fermionic-like in some others regions of the $\{\varepsilon_0,eV\}$ plane. Thus, a positive sign in the real part of the cross-correlator can be seen as the seal of the 
Coulomb interactions present in the QD.

We now focus on the effect of interactions on the profile of the auto-correlators $\mathcal{S}_{LL}(\nu)$ and  $\mathcal{S}_{RR}(\nu)$ shown on the first and second columns in Fig.~\ref{figure}. We remark that the intensity of the auto-correlators is reduced 
when interactions are present (compare the color scale intensities in the graphs of Figs.~\ref{figure}(c) and \ref{figure}(d) to the ones of Figs.~\ref{figure}(a) and \ref{figure}(b)), in full agreement with the fact that the charge becomes frozen when the QD is in the 
Kondo regime\cite{Sela2006,Desjardins2017}, leading to a reduction of the noise. We also remark the doubling of the number of red triangles in the color-scale plots of $\mathcal{S}_{LL}(\nu)$ and  $\mathcal{S}_{RR}(\nu)$ and the appearance of a more 
complex structure when $U\ne 0$ in comparison to the $U=0$ case: notably, there appears a Coulomb diamond-like structure, centered around the point of coordinates $(\varepsilon_0=-U/2,eV=0)$ in the $\{\varepsilon_0,eV\}$ plane, inside which the noise is 
strongly reduced. This means that by setting adequately the values of $\varepsilon_0$ and $eV$ inside the region defined by this structure, one could reduce drastically the noise in experiments.

Finally, we discuss the effect of the coupling asymmetry on the noise color-scale plots. Whereas the dependences of the auto-correlators $\mathcal{S}_{LL}(\nu)$ and  $\mathcal{S}_{RR}(\nu)$ are neither odd nor even functions of both $\varepsilon_0$ and $eV
$, we note that the real part of the cross-correlator $\mathcal{S}_{LR}(\nu)$ and the total noise $\mathcal{S}_\mathrm{tot}(\nu)$ are even functions of both $\varepsilon_0$ and $eV$ when the couplings are symmetrical ($a=1$) as shown in Figs.~\ref{figure}(a) 
and \ref{figure}(c). This is no longer the case for asymmetric couplings ($a=4$) as shown in Figs.~\ref{figure}(b) and \ref{figure}(d). We also observe that in the presence of interactions the noise is strengthened in the less-coupled reservoir, here the $R$ 
reservoir since the value $a=4$ corresponds to $\Gamma_R=\Gamma_L/4$. Intuitively, this happens because the transmission of carriers from the $R$ reservoir to the QD is weaker, and it is this transmission which mainly contributes to $\mathcal{S}_{LL}(\nu)$, 
than the transmission from the $L$ reservoir to the QD, which mainly contributes to $\mathcal{S}_{RR}(\nu)$.

%%%%%%%%%%%%%%%%%%%%%%%%%%%%%%%%%%%%%%%%%%%%%%%%%%%%%%%%%%%%%%%%%%
%																 %
%																 %
%		CONCLUSION											 %
%																 %
%																 %
%%%%%%%%%%%%%%%%%%%%%%%%%%%%%%%%%%%%%%%%%%%%%%%%%%%%%%%%%%%%%%%%%%

\section{Conclusion}
We have developed a theory to calculate the finite-frequency noise in a non-equilibrium Kondo QD, which allows us to analyze the features observed in the evolution of the noise as a function of dot energy level and bias voltage. We have discussed the effect of 
the asymmetry in the couplings to the reservoirs. We predicted a change of sign in the real part of the cross-correlator when interactions are present in the QD; this is related to the fact that the statistics of the carriers are no longer fermionic. We also highlighted 
the appearance at $U\ne 0$ of a Coulomb diamond like structure in the auto-correlators and total noise profiles inside which the fluctuations are reduced.

{\it Acknowledgments} -- The authors want to thank H.~Baranger, H.~Bouchiat, R.~Deblock, R.~Delagrange, M.~Guigou, F.~Michelini and X.~Waintal for valuable discussions. For financial support, the authors acknowledge the Indo-French Center for the 
Promotion of Advanced Research (IFCPAR) under Research Project No.4704-02.

%%%%%%%%%%%%%%%%%%%%%%%%%%%%%%%%%%%%%%%%%%%%%%%%%%%%%%%%%%%%%%%%%%
%																 %
%																 %
%		REFERENCES											     %
%																 %
%																 %
%%%%%%%%%%%%%%%%%%%%%%%%%%%%%%%%%%%%%%%%%%%%%%%%%%%%%%%%%%%%%%%%%%

\end{document}